\PassOptionsToPackage{unicode}{hyperref}
\PassOptionsToPackage{hyphens}{url}
\documentclass[12pt]{article}
\usepackage[margin=1in]{geometry}  
\usepackage{amsmath,amssymb}
\usepackage{lmodern}
\usepackage{graphicx}
\usepackage{hyperref}
\usepackage{natbib}
\usepackage{xcolor}
\usepackage{booktabs}
\usepackage{longtable}
\hypersetup{
  colorlinks=true,
  linkcolor=black,
  urlcolor=blue,
  citecolor=black
}
\ifPDFTeX
  \usepackage[T1]{fontenc}
  \usepackage[utf8]{inputenc}
  \usepackage{textcomp} 
\else 
  \usepackage{unicode-math}
  \defaultfontfeatures{Scale=MatchLowercase}
  \defaultfontfeatures[\rmfamily]{Ligatures=TeX,Scale=1}
\fi
\IfFileExists{upquote.sty}{\usepackage{upquote}}{}
\IfFileExists{microtype.sty}{
  \usepackage[]{microtype}
  \UseMicrotypeSet[protrusion]{basicmath} 
}{}
\makeatletter
\@ifundefined{KOMAClassName}{
  \IfFileExists{parskip.sty}{%
    \usepackage{parskip}
  }{
    \setlength{\parindent}{0pt}
    \setlength{\parskip}{6pt plus 2pt minus 1pt}}
}{
  \KOMAoptions{parskip=half}}
\makeatother
\usepackage{xcolor}
\usepackage{longtable,booktabs,array}
\usepackage{calc} 
\usepackage{etoolbox}
\makeatletter
\patchcmd\longtable{\par}{\if@noskipsec\mbox{}\fi\par}{}{}
\makeatother
\IfFileExists{footnotehyper.sty}{\usepackage{footnotehyper}}{\usepackage{footnote}}
\makesavenoteenv{longtable}
\usepackage{graphicx}
\makeatletter
\def\maxwidth{\ifdim\Gin@nat@width>\linewidth\linewidth\else\Gin@nat@width\fi}
\def\maxheight{\ifdim\Gin@nat@height>\textheight\textheight\else\Gin@nat@height\fi}
\makeatother
\setkeys{Gin}{width=\maxwidth,height=\maxheight,keepaspectratio}
\makeatletter
\def\fps@figure{htbp}
\makeatother
\usepackage[normalem]{ulem}
\ifLuaTeX
  \usepackage{selnolig}  
\fi
\IfFileExists{bookmark.sty}{\usepackage{bookmark}}{\usepackage{hyperref}}
\IfFileExists{xurl.sty}{\usepackage{xurl}}{} 
\hypersetup{
  hidelinks,
  pdfcreator={LaTeX via pandoc}}

\author{}
\date{}

\begin{document}

\title{Beyond Misinformation: A Conceptual Framework for Studying AI Hallucinations in (Science) Communication}
\author{Anqi Shao\\
 \\
Department of Life Sciences Communication\\
University of Wisconsin–Madison\\
anqi.shao@wisc.edu}
\date{April 2025}

\maketitle

\begin{abstract}
This paper proposes a conceptual framework for understanding AI hallucinations as a distinct form of misinformation. While misinformation scholarship has traditionally focused on human intent, generative AI systems now produce false yet plausible outputs absent of such intent. I argue that these AI hallucinations should not be treated merely as technical failures but as communication phenomena with social consequences. Drawing on a supply-and-demand model and the concept of distributed agency, the framework outlines how hallucinations differ from human-generated misinformation in production, perception, and institutional response. I conclude by outlining a research agenda for communication scholars to investigate the emergence, dissemination, and audience reception of hallucinated content, with attention to macro (institutional), meso (group), and micro (individual) levels. This work urges communication researchers to rethink the boundaries of misinformation theory in light of probabilistic, non-human actors increasingly embedded in knowledge production.
\end{abstract}


\newpage

\section{Introduction}
In January 2024, an AI-generated robocall impersonating U.S. President
Joe Biden urged New Hampshire voters to avoid voting in the Democratic
primary (Kaczynski, 2024). The call, later traced to a political
consultant using generative AI tools, was condemned as an election
interference attempt (Shepardson, 2024). This is not an isolated case
but is part of the growing use of generative AI to amplify and automate
misinformation production, making deception easier, cheaper, and more
scalable (Bond, 2024). While traditional misinformation typically stems
from human oversights or fabrication (intentional or unintentional
falsehoods), these cases represent a shift toward misinformation
co-produced by humans and AI.

Such AI-generated misinformation in political campaigns is often
associated with direct human manipulation. However, AI systems powered
by large language models (LLMs) can also produce inaccuracies without
deliberate human intent. Unlike human agents who may actively search,
verify, and summarize information, AI generates responses
probabilistically. It constructs text based on statistical patterns
learned from vast datasets rather than a direct understanding of factual
accuracy, which makes these systems susceptible to hallucinations: false
or misleading outputs that appear plausible but lack factual grounding.
Despite this limitation, AI is now widely embedded in processes of
public knowledge formation such as online search, customer service ,
journalism, and scientific research (Reid, 2024). At least 46\% of
Americans reported that they have used AI tools for information seeking
(IPSOS, 2025), though this figure might be underestimated -\/- a recent
survey showed that while 99\% of Americans had used a product with AI
features yet only 64\% recognized they have ever used one (Gallup,
2024). Users may unknowingly rely on AI generated content and assume it
functions like a traditional information source when in reality it
produces output based on statistical probabilities rather than
independently verified facts.

The real-world consequences of these AI hallucinations are already
evident across various domains. Concerns have emerged in medical
applications, where OpenAI\textquotesingle s Whisper system, a
transcription AI, misrepresented user speech and fabricated false or
misleading content in its hallucinated transcriptions (Koenecke et al.,
2024). In business, hallucinations have led to operational and legal
issues: for instance, Air Canada's chatbot provided misleading
information about bereavement fares, resulting in a misinformed customer
and associated financial loss (\emph{Moffatt v. Air Canada}, 2024). In
academia, LLMs have demonstrated notable failures, creating non-existing
citations in a legal declaration due to unverified ChatGPT outputs
(Zhao, 2024) and making systematic errors in academic terms when editing
manuscripts (Oransky, 2024). AI's application in established news
agencies has faced similar challenges, such as publishing AI-generated
content that contained misleading historical claims (Owen, 2025; Reilly,
2025). Beyond direct usage of AI, the Google search engine recently
cited an April Fool's satire as fact in its AI summary of search
results, falsely claiming microscopic bees were used in computing
(Kidman, 2025). Further studies suggest that such errors are not
anomalies but statistically inevitable. The estimated chance of AI
hallucination is subject to a statistical lower bound (Kalai \& Vempala,
2024), with earlier studies reporting rates ranging from 5\% for general
queries to 29\% for specialized professional questions in benchmark
testing (Lukens \& Ali, 2023). More recent research suggests the
hallucination rates may currently be between 1.3\% and 4.1\%, in tasks
such as text summarization (Vectara, 2024).

The unavoidable nature of AI hallucinations, demonstrated by these
examples and studies, makes them distinct from conventional of
misinformation with controllable human intent. This lack of clear intent
challenges traditional frameworks, as Schäfer (2023) notes: ``One
theoretical gap lies in how we conceptualize misinformation when AI is
involved: traditional distinctions between misinformation and
disinformation blur if a non-human agent produces the falsehood without
clear intent.'' This distinction places AI hallucinations at the
boundary of what scholars define as misinformation: Are they merely
technical errors, or do they function similarly to human-generated
misinformation in shaping public perceptions and decision-making? This
essay argues that hallucinations produced by generative AI tools,
powered by LLMs, constitute a distinct form of misinformation and should
be analyzed as such. I offer a conceptual framework for both
understanding and studying AI hallucinations as a new and distinct
category of misinformation. They emerge from the probabilistic nature
and socio-technical deployment of Ais, rather than directly from human
oversight or deliberate intent. Based on this framework, I outline a
research agenda for communication scholars, focusing on how
hallucinations are generated (supply), how they are interpreted and
shared (demand), and what makes them persuasive in public discourse.

\hypertarget{beyond-misinformation-a-conceptual-framework-for-studying-ai-hallucinations-as-a-new-source-of-inaccuracy}{%
\subsection{Beyond Misinformation: a Conceptual Framework for Studying
AI Hallucinations as a New Source of
Inaccuracy}\label{beyond-misinformation-a-conceptual-framework-for-studying-ai-hallucinations-as-a-new-source-of-inaccuracy}}

Misinformation is not a new phenomenon; it is an old wine constantly
rebranded in new bottles for specific agendas. During the COVID
pandemic, the World Health Organization (WHO) warned of an ``infodemic''
that overwhelmed public audiences with a deluge of informational noise,
and made it difficult for citizens to distinguish accurate information
from irrelevant, false, or misleading content (World Health
Organization, 2020). U.S. Surgeon General Dr. Vivek Murthy echoed this
concern and has identified health misinformation as an ``urgent threat''
and stated that it prolonged the COVID-19 pandemic and endangered lives
through misinformation-fueled behaviors (Brumfiel, 2021).

Rather than viewing misinformation solely as a crisis to be solved,
however, it is more productive to examine why misinformation emerges,
and how it operates in the current information ecosystem and influences
public reasoning. Framing misinformation as a unique crisis may neglect
it as a persistent part of democratic discourse, which involved
competing claims, and misinformation is often a byproduct of evolving
knowledge, ideological conflicts, and contested narratives within
political and scientific deliberations (Budak et al., 2024; Krause et
al., 2022; Scheufele, Krause, et al., 2021). Take science communication
as an example -\/- The scientific process is inherently provisional,
with findings continually subject to revision. Early results frequently
evolve or become overturned by subsequent research and rigorous peer
review. Consequently, preliminary or controversial findings can
inadvertently lead to misinformation if prematurely communicated as
definitive facts (National Academies of Sciences, Engineering, and
Medicine, 2024). During the COVID-19 pandemic, for instance, the rapid
proliferation of non-peer-reviewed preprints demonstrated how scientific
uncertainty could amplify misinformation, as initial results were
sometimes misinterpreted or misrepresented as conclusive evidence
(Krause et al., 2022). Additionally, systemic incentives in science
communication often encourage exaggerated claims, hype, or
sensationalized interpretations, especially regarding emerging
technologies such as artificial intelligence or groundbreaking
scientific discoveries (Scheufele, Hoffman, et al., 2021). Even after
correction through peer review, these sensationalized narratives can
persist in public memory, further perpetuating misinformation. Attempts
to treat misinformation primarily as a problem to be ``solved'' risk
falling into the knowledge deficit model, which assumes that simply
replacing false information with accurate facts would lead to better
decision-making in democratic societies (Akin, 2017; Scheufele, 2014).

Here, my working definition of misinformation is broadly understood as
any content that contradicts the best available evidence (Scheufele \&
Krause, 2019). This definition covers both unintentional errors and
deliberate deceptions, making it distinct from disinformation, which
refers to intentional falsehoods designed to mislead. To analyze the
dynamics of misinformation within contemporary information ecosystems, I
apply a supply-and-demand framework for existing theoretical and
conceptual constructions of human-initiated misinformation. This
approach divides the study of misinformation into two primary aspects:
-\/- the origins and propagation of misinformation, and (2) the demand
side -\/- its consumption and dissemination by the public. On the supply
side, human-generated misinformation often arises from epistemic
limitations like preliminary or unsettled scientific results,
publication biases favoring sensational outcomes, and cognitive biases
within the scientific community itself, reinforced by peer-review echo
chambers (Scheufele, 2014; Scheufele, Hoffman, et al., 2021). On the
demand side, misinformation persists due to public cognitive biases,
heuristic reasoning, motivated reasoning, and varying degrees of trust
in scientific institutions (Hart \& Nisbet, 2012; Schäfer, 2016).

\emph{\textbf{AI Hallucinations are \uline{Technically} Different from
Human Misinformation}}

While this supply-and-demand framework offers valuable insights into the
persistent challenges of human-generated misinformation -\/- areas where
considerable research exists -\/- the emergence of new communication
technologies demands an expanded perspective. Specifically, generative
AI powered by LLMs introduces novel mechanisms for generating
falsehoods, such as hallucinations, which function differently from
human-driven misinformation. Although these AI outputs can similarly
pollute the information ecosystem, their underlying causes and
characteristics require distinct analysis. To bridge our understanding
of established misinformation dynamics with this emerging challenge,
\textbf{Table 1} summarizes and compares key differences between
human-generated misinformation and AI-generated hallucinations. This
comparison underscores the unique nature of AI hallucinations and
highlights the need for focused investigation, which will be the subject
of the following sections.

\textbf{Table 1:} Comparing misinformation challenges in human vs. AI
contexts.

\begin{longtable}[]{@{}
  >{\raggedright\arraybackslash}p{(\columnwidth - 4\tabcolsep) * \real{0.1730}}
  >{\raggedright\arraybackslash}p{(\columnwidth - 4\tabcolsep) * \real{0.4245}}
  >{\raggedright\arraybackslash}p{(\columnwidth - 4\tabcolsep) * \real{0.4025}}@{}}
\toprule()
\begin{minipage}[b]{\linewidth}\raggedright
\end{minipage} & \begin{minipage}[b]{\linewidth}\raggedright
Human
\end{minipage} & \begin{minipage}[b]{\linewidth}\raggedright
AI
\end{minipage} \\
\midrule()
\endhead
Supply & Limited by cognitive and epistemic boundaries (settled,
knowable, and unsettled science). & Lacks genuine understanding; unaware
of its knowledge limits. \\
Demand & Driven by accuracy, biases, or other motivations; debates on
efficacy of accuracy priming. & Errors arise from probabilistic modeling
and algorithmic limitations. \\
\bottomrule()
\end{longtable}

AI hallucinations, distinct from the psychological phenomena experienced
by humans, stem from the inherent design and operational principles of
large language models (LLMs). These systems sometimes generate outputs
that deviate from the user\textquotesingle s input, contradict previous
context, or fail to align with established facts---a phenomenon broadly
termed ``hallucination'' (Zhang et al., 2023). Within computer science,
AI hallucinations are often categorized into two types: faithfulness
errors, where the generated text does not accurately reflect the
provided source or input context, and factualness errors, where the
generated text misaligns with established real-world knowledge (Ji et
al., 2023, Augenstein et al., 2024).

Understanding why LLMs produce hallucinations requires examining
multiple contributing factors inherent to their design and deployment.
These factors can be conceptualized using a layered risk model, akin to
the Swiss cheese model often used in risk analysis (see \textbf{Figure
1}), where vulnerabilities at different stages can align to allow errors
to manifest. Key contributing layers include the nature of the training
data, the process of text generation, and limitations in downstream
gatekeeping

\textbf{Figure 1:} The Swiss cheese model of the vulnerabilities that
causes AI hallucination

\includegraphics[width=4.83978in,height=2.81239in]{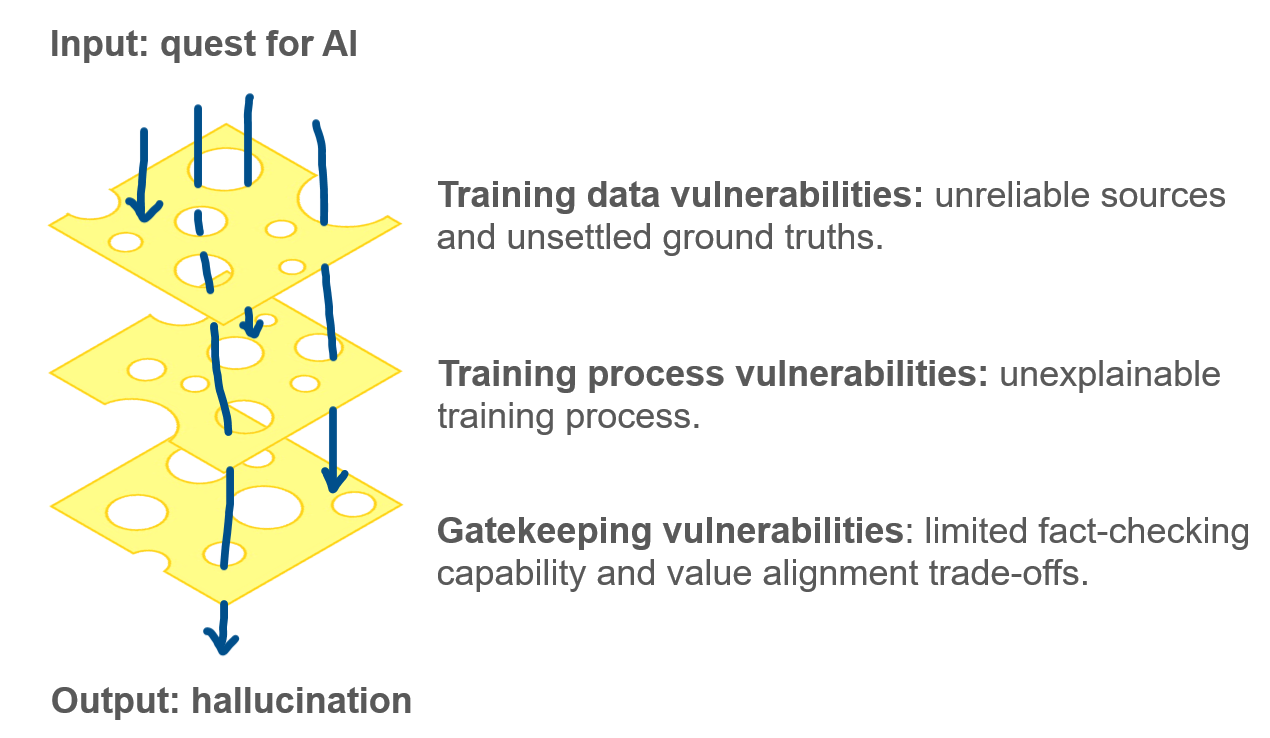}

First, vulnerabilities exist within the training data itself. The
training datasets used to build LLMs often reflect the biases, errors,
and inconsistencies of the humans who create them (Chen et al., 2024).
These flaws can propagate into the models, embedding systemic issues
directly into their outputs (Xu et al., 2025). Additionally,
inaccuracies or omissions in the training data can lead to
hallucinations when the model attempts to generate content beyond the
scope of its learned information (Tian et al., 2023) and even ``model
collapse'' due to the lack of fresh human-generated data (Shumailov et
al., 2024). While techniques like retrieval-augmented generation (RAG)
aim to mitigate knowledge gaps by providing external, up-to-date
information (Fan et al., 2024), they do not eliminate the risk.
Challenges remain, such as (a) the potential unreliability of retrieved
sources (e.g., poisoned RAG in Zou et al., 2024), echoing the core data
quality problem, and (b) difficulty resolving conflicting information,
particularly for unsettled or ambiguous topics where no ground truth
exists (Scheufele, Hoffman, et al., 2021). Thus, claims that
hallucinations can be completely eliminated -\/- often requiring
unrealistic conditions like perfectly structured and clean input -\/-
remain largely aspirational (e.g., Wood \& Forbes, 2024).

Second, hallucinations are rooted in the probabilistic nature of LLM
text generation. These models function by predicting the most
statistically likely next word (or token) in a sequence, based on
patterns learned from their training data (Xu et al., 2025). While this
approach enables the generation of coherent and contextually appropriate
language, it does not guarantee factual accuracy (Zhang et al., 2023).
LLMs estimate the likelihood of the next word based on the preceding
context. If the learned probability distribution is biased, incomplete,
or too general, the model might produce outputs that are statistically
probable but factually incorrect (Xu et al., 2025). Furthermore,
parametric knowledge limitations mean the information implicitly stored
within the model\textquotesingle s parameters is confined to its
training data, hindering accurate generalization to novel scenarios and
sometimes leading to overconfident yet inaccurate responses
(Thirunavukarasu et al., 2023). In light of the reasons for
hallucination above, the probabilistic nature of LLMs makes
hallucinations an inherent limitation rather than an occasional glitch.

Finally, a third layer of risk involves flaws in gatekeeping mechanisms
designed to catch errors before outputs reach users. Current approaches
struggle to detect and prevent all hallucinations due to several
factors: the sheer volume of AI-generated content makes comprehensive
human review infeasible (Montasari, 2024); subtle hallucinations, such
as slightly incorrect figures or plausible but fabricated references,
are difficult to identify (Zhao, 2024); and automated fact-checking
systems are imperfect and may miss context-dependent or domain-specific
errors (Lu, 2025).

These inherent characteristics position AI hallucinations uniquely as a
source of inaccurate information. While lacking the clear human intent
often associated with disinformation, their capacity to appear plausible
and influence user understanding raises critical questions. Do they
function similarly to traditional misinformation in shaping perceptions
and decisions, despite their different origins? This functional impact,
combined with their statistical inevitability and the blurred lines
around algorithmic \textquotesingle intent,\textquotesingle{}
necessitates treating AI hallucinations not just as technical errors,
but as a distinct category of misinformation requiring dedicated
conceptualization and research within communication studies.

\hypertarget{ai-hallucinations-are-conceptually-different-from-human-misinformation}{%
\subsubsection{\texorpdfstring{AI Hallucinations are
\uline{Conceptually} Different from Human
Misinformation}{AI Hallucinations are Conceptually Different from Human Misinformation}}\label{ai-hallucinations-are-conceptually-different-from-human-misinformation}}

Traditional frameworks for misinformation often center solely on human
actors, focusing on their knowledge states and intentions. However, this
approach is insufficient for explaining AI-generated inaccuracies, which
arise from interactions between humans and AI systems. Understanding
these inaccuracies requires acknowledging distributed agency, where
agency (the capacity to act and produce effects) is not located within a
single entity but emerges from relationships and interactions within a
network of heterogeneous actors -\/- both human and non-human (e.g.,
algorithms, datasets, interfaces) (Rammert, 2008). Actor-Network Theory
(ANT), for instance, suggests that actions and outcomes are generated
through the associations formed within such networks (e.g. Latour,
2005).

\textbf{Figure 2} visually models this concept of distributed agency
using a circular spectrum to illustrate the continuum of human
involvement in the production of inaccurate information. The shading
represents the degree of human engagement and control in the process,
ranging from high (darker shading) to low (lighter shading). The
spectrum highlights how different forms of inaccuracy emerge based on
the nature and intensity of human interaction with AI systems.

\textbf{Figure 2:} A ring of inaccuracies: The distributed agency of
AI-infused misinformation production (Hallucination vs. Human-initiated;
The darker the shading, the stronger human agency engages with the
process).

\includegraphics[width=5.14583in,height=3.04497in]{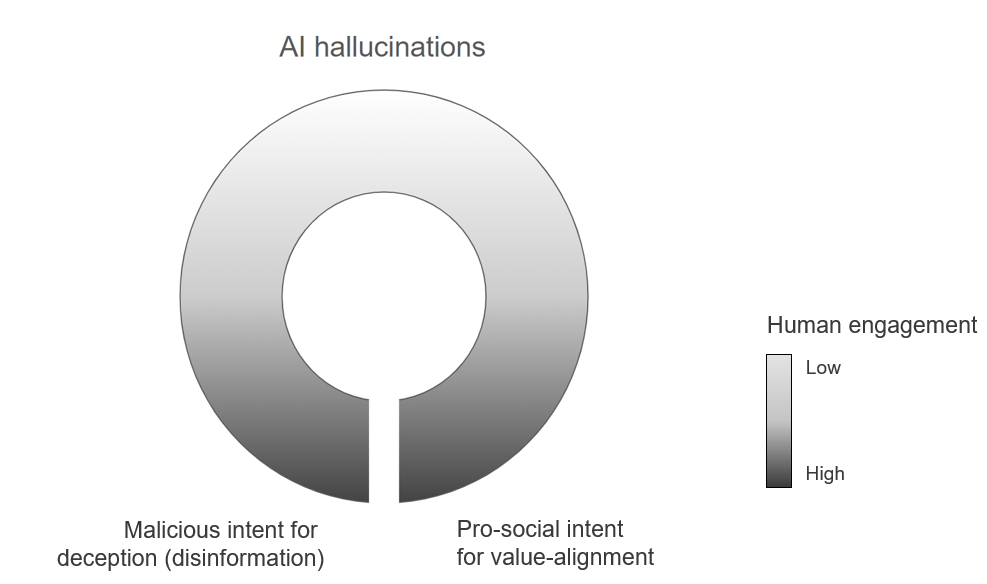}

The bottom arc of Figure 2 represents scenarios with high human
engagement and control. On the right side (Human-Initiated
Inaccuracies), humans exert significant control in generating
falsehoods. This includes deliberate disinformation, such as using AI
tools to create deepfakes with malicious intent (e.g., the malicious use
of deepfake, Westerlund, 2019) but also unintentional inaccuracies
resulting from humans directing AI with flawed prompts or biased inputs
(Zamfirescu-Pereira et al., 2023). In these cases, the primary locus of
agency and intentionality (or lack thereof, in the case of errors)
resides with the human actor, even if AI is used as a tool. On the left
side (Human-Influenced Inaccuracies), human control is exercised more
indirectly through the imposition of constraints, guardrails, or
value-alignment objectives during AI development or deployment. While
the intent might be pro-social (e.g., mitigating harmful stereotypes),
these human-imposed values can lead the AI to produce outputs that
distort factual reality, such as misrepresenting demographic
distributions in generated images (Thorbecke \& Duffy, 2024) . Here,
inaccuracy is a byproduct of human choices influencing the
AI\textquotesingle s operational boundaries.

Conversely, the top arc of the spectrum represents AI Hallucinations,
characterized by minimal direct human control or engagement in the
specific generation of inaccuracies. As detailed previously,
hallucinations stem primarily from the inherent operational
characteristics of LLMs -\/- their probabilistic nature, data
limitations, and imperfect optimization -\/- rather than specific human
directives or intentions at the point of generation. While human prompts
initiate the process, the specific content of a hallucination is largely
an emergent property of the AI system itself. Moving slightly rightward
from the top suggests some human influence via prompts potentially
leading the AI into hallucination-prone territory, while moving slightly
leftward suggests human-imposed guardrails subtly shaping the output
landscape where hallucinations might still occur; in both cases, direct
human control over the specific inaccuracy remains low.

These different ways inaccuracies are generated, shown in \textbf{Figure
2}, align with the contrasts drawn earlier between human misinformation
and AI outputs (see \textbf{Table 1}). In essence, AI hallucinations
arise differently because LLMs operate distinctively from humans: they
produce text based on statistical patterns without the understanding or
awareness of knowledge limits that bound human communication (Xu et al.,
2024; Tian et al., 2024). Furthermore, the inaccuracies produced
(hallucinations) result from the AI\textquotesingle s probabilistic
methods and technical limitations, not from psychological motivations or
communicative intent driving human errors or deception. This absence of
human-like knowledge processing and intention behind the generation of
falsehood itself is the core theoretical distinction.

These specific operational characteristics make AI hallucinations a
distinct source of inaccurate information. While lacking clear human
intent is often associated with disinformation, their capacity to appear
plausible and influence user understanding raises new questions for
researchers who may want to shift attention from human misinformation to
AI hallucinations. Do they function similarly to traditional
misinformation in shaping perceptions and decisions, despite their
different origins? This functional impact, combined with their
statistical nature and the lack of clear algorithmic ``intent,''
suggests AI hallucinations should be treated not just as technical
errors, but as a distinct category of misinformation requiring
conceptualization and research within communication studies.

In summary, viewing AI-related inaccuracies through the lens of
distributed agency, as illustrated in \textbf{Figure 2}, allows for a
clearer understanding than traditional human-centric models permit. This
framework clarifies how different types of falsehoods emerge from
varying configurations of human control, engagement, and AI behavior. It
highlights the unique theoretical position of AI hallucinations, showing
they stem from different production mechanisms compared to human
misinformation, occurring with minimal direct human control and no
psychological intentionality.

\hypertarget{where-we-go-next-a-research-agenda-for-ai-hallucination-in-communication}{%
\subsection{Where We Go Next: A Research Agenda for AI Hallucination in
Communication}\label{where-we-go-next-a-research-agenda-for-ai-hallucination-in-communication}}

AI hallucinations are not just technical flaws. While existing
communication frameworks emphasize intention, belief, and cognition in
human misinformation, hallucinations arise from systems that generate
plausible yet inaccurate content without intent or awareness. To treat
them merely as a subset of misinformation risks overlooking what makes
them distinct. This section moves beyond simply identifying the problem
and instead outlines key questions and priorities for understanding and
addressing the impact of AI hallucinations. Building on the
supply-and-demand perspective used for traditional misinformation, this
agenda identifies key questions and directions needed to understand how
AI hallucinations are produced, circulated, and impact individuals and
society

\hypertarget{not-just-a-bug-an-agenda-for-addressing-the-supply-side-of-ai-hallucinations}{%
\subsubsection{\texorpdfstring{Not just a bug: an agenda for addressing
the \uline{supply} side of AI
hallucinations}{Not just a bug: an agenda for addressing the supply side of AI hallucinations}}\label{not-just-a-bug-an-agenda-for-addressing-the-supply-side-of-ai-hallucinations}}

Understanding the ``supply'' of AI hallucinations involves examining the
upstream factors that contribute to their generation and propagation
before they reach audiences, often without any deceptive intent. Key
areas may include:

\textbf{Knowledge Boundaries \& Uncertainty.} AI systems often provide
answers even when dealing with unsettled science or topics lacking clear
or credible ground truths (Augenstein et al., 2024; Kington et al.,
2021). While scientists themselves grapple with communicating the
uncertainties of their work, as media often simplify their findings for
wider public consumption (Beets, 2024; Dunwoody et al., 2018; Peters \&
Dunwoody, 2016), it remains unclear how AI models will deal with such
uncertainty. For instance, it remains unknown how the reliance on
hedging or disclaimers (e.g., the mini-sized ``XXX can make mistakes.
Check important info'' disclaimer under the interface of popular
generative AI tool interfaces) of AI-generated contents would have on
user perceptions. Furthermore, even in answering questions where
scientific consensus exists, AI systems can falter. Their use of
metaphor and simplification, meant to aid understanding, can also
mislead (e.g., in \textbf{Figure 3} below, where ChatGPT boldly suggests
such prompts for users about scientific ideas). For instance, in
human-lead science communication, referring to anticoagulants as ``blood
thinners'' is a normatively accepted simplification among physicians,
yet technically inaccurate (National Academies of Sciences, Engineering,
and Medicine, 2024). Given the precedent of human simplification leading
to misinformation, a parallel key question is whether AI's
simplifications have similar normative value, or whether they cross into
distortion that shapes public reasoning in unintended ways.

\textbf{Figure 3}: A Screenshot from the ChatGPT mobile APP landing page
in January 2025.

\includegraphics[width=2.34124in,height=2.4271in]{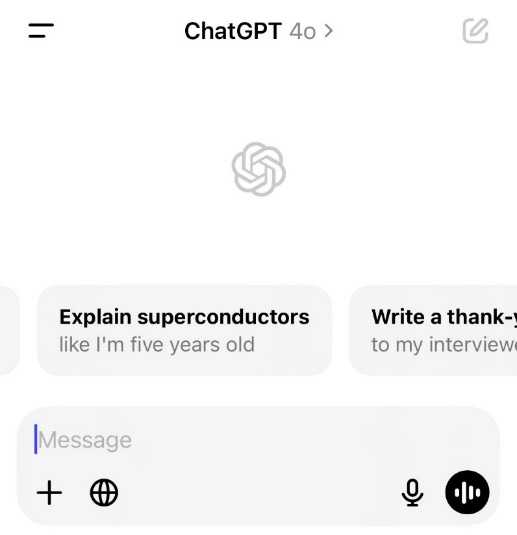}

\textbf{Data Logistics \& Biases.} The data used to train LLMs suffer
from limitations like gaps (``data voids'', Golebiewski \& Boyd, 2019),
biases reflecting societal inequalities (Crawford, 2021; ``digital
universalism'', Loukissas, 2019), and quality issues (Wood \& Forbes,
2024). These problems can lead to biased or inaccurate outputs,
disproportionately affecting certain user groups (Chen et al., 2024),
and the privacy regulations (e.g., General Data Protection Regulation or
GDPR, Voigt \& von dem Bussche, 2017) and platform opacity (Brennen et
al., 2025) hinders research into these effects. This lack of
transparency means that much of our understanding is based on anecdotes
or small samples (Krause et al., 2024). As a result, many claims about
AI in communications (either positive or negative) remain unexplained.
Communication studies can document how these data issues translate into
information inequities when AI interacts with diverse user groups (e.g.,
Chen et al., 2024). The potential homogenization of information due to
limited data diversity (Tewari et al., 2021) also warrants attention
from those studying media ecosystems to avoid providing anecdotal
evidence.

\textbf{Opacity of AI Processes.} The internal workings of LLMs remain
largely opaque or ``black boxes,'' making it difficult to pinpoint why
specific hallucinations occur (Bender et al., 2021; Weidinger et al.,
2021; Zhang et al., 2023). Tools in explainable AI (XAI) offer partial
insight, but they do not resolve the deeper problem: hallucinations
emerge from interactions between opaque data, training choices, and user
inputs. Even small changes in fine-tuning -\/- such as optimizing a
model for one domain -\/- can create unanticipated distortions elsewhere
(Betley et al., 2025). Instead of one-off creating fake audio or video
(like a deepfake), it is a systematic effort to poison the model's
supply of training data or tasks.

From a communication perspective, this shifts the question from
\emph{``what is true?''} to \emph{``why does the system produce what it
does, and how will the information be interpreted?''} Communication
scholars have historically addressed the social construction of
knowledge and misinformation by identifying interventions along the
message production and dissemination process (e.g., Lewandowsky et al.,
2012; Pennycook et al., 2021). However, a parallel understanding of the
internal mechanisms driving LLM hallucinations is lacking. Intervention
cannot focus solely on model behavior. It also involves design choices
around interfaces, disclosures, and feedback mechanisms. Therefore,
insights from communication research on user perception should be
combined with Human-Computer Interaction (HCI) expertise (Schäfer,
2023). HCI provides methods for designing and evaluating interfaces to
make AI systems more understandable and transparent from a
human-centered viewpoint (e.g., Liao \& Vaughan, 2024) and for
developing interaction techniques that help people better anticipate or
manage AI errors, thereby improving human-AI collaboration (e.g., Ozmen
Garibay et al., 2023)

\textbf{Gatekeeping \& Alignment Trade-offs.} Hallucinations expose the
limits of traditional gatekeeping. Fact-checking, whether through
experts or automated tools, is labor-intensive and often retrospective
(e.g., Omiye et al., 2023; Palta et al., 2024). Subtle inaccuracies,
such as fabricated citations or plausible misstatements, may pass
unnoticed. Another challenge in this step is alignment: the attempt to
shape LLM outputs according to socially acceptable norms. Despite the
growing interest in aligning LLMs with human values (e.g., Kirk et al.,
2024), such alignment often risks distorting the reality these systems
aim to represent -\/- that is, enforcing certain human values may lead
to outputs that prioritize social acceptability at the expense of
factual accuracy. For instance, AI image generators have produced
historically implausible depictions under the banner of diversity,
prompting public backlash (Thorbecke \& Duffy, 2024). These are not
simple errors but outcomes of designers' choices --- decisions about
what \emph{should be} shown or said, not necessarily what \emph{it is}.

Institutional responses, such as banning ChatGPT from co-authorship or
issuing guidelines for AI use in publishing (Haggart, 2023) are rather
reactive. Because hallucinations are difficult to anticipate, policies
struggle to address their emergence in real time. Given that
AI-generated content and its inaccuracies cannot be reliably forecasted,
institutions may encounter ongoing uncertainty in regulatory processes
to unexpected manifestations of AI behavior. Given emerging suggestions
that domain-specific tolerance for AI errors varies (Lu, 2025),
communication research can play a role beyond ``hallucination
catchers''. A possible area for study involves how norms about
``acceptable error'' is defined and negotiated, and how these norms vary
across audiences, domains, and platforms. Understanding these social
dynamics is distinct from the technical task of error detection.

\hypertarget{more-persuasive-than-misinformation-an-agenda-for-studying-the-demand-side-of-hallucinations}{%
\subsubsection{\texorpdfstring{More persuasive than misinformation? An
agenda for studying the \uline{demand} side of
hallucinations}{More persuasive than misinformation? An agenda for studying the demand side of hallucinations}}\label{more-persuasive-than-misinformation-an-agenda-for-studying-the-demand-side-of-hallucinations}}

On the demand side, the central question is: what attributes make AI
hallucinations persuasive enough for audiences to accept and believe
them? Unlike traditional misinformation, AI hallucinations often present
themselves without intentional manipulation but are perceived as
credible due to their coherence, clarity, and authoritative tone (Zhang
et al., 2023). Existing research ``tend{[}s{]} to study isolated
phenomena confined within a single platform'' but must instead be
situated ``into larger systems of (mediated) communication and
interaction'' (Krause et al., 2024). Communication in today's
multi-modal information ecologies cannot be studied in isolation from
broader social and institutional contexts. In this section, I propose a
macro--meso--micro framework for analyzing how AI hallucinations
function within these layered systems.

\textbf{Macro-Level: Institutional Roles and Media Credibility.}
Traditional mechanisms for ensuring institutional credibility of
information, such as fact-checking and source evaluation, often rely on
identifying human intentions behind misinformation. However, the
probabilistic and often unintentionally misleading nature of AI outputs
challenge traditional verification processes. While interventions
developed for human misinformation, such as accuracy nudges encouraging
reflection before sharing (e.g., Pennycook et al., 2021), psychological
inoculation preemptively building resistance (van der Linden, 2023), or
more recently, leveraging AI tools in providing indefatigable
interventions towards conspiracists (Costello et al., 2024), offer
potential starting points, their applicability and effectiveness against
unintentionally inaccurate AI content require specific examination.
Consequently, adapting verification methodologies specifically for
AI-generated content also becomes warranted, potentially incorporating
computational fact-checking and human-AI hybrid methods that are
currently used (Narayanan Venkit et al., 2024), or appropriately
modified psychologically informed strategies to reliably detect and
mitigate inaccuracies. Moreover, media credibility in the age of AI
faces distinct challenges. While disclosure of AI-generated content has
shown mixed results regarding perceived accuracy (Bien-Aimé et al.,
2025; Li et al., 2025), the possibility of trust if AI involvement is
recognized suggests a need for new institutional strategies for
transparency and credibility maintenance.

\textbf{Meso-Level: Group Dynamics and (Online) Dissemination.} The
meso-level refers to the group contexts in which information spreads
through interactions and social networks. At this level, prior
discussions on how human misinformation is received and disseminated
have focused on phenomena such as echo chambers, where individuals
primarily encounter information reinforcing their existing views
(Jamieson \& Cappella, 2008), filter bubbles, resulting from
algorithmically personalized information spaces that restrict content
variety (Cinelli et al., 2021), and motivated reasoning, which refers to
the tendency to process information in ways that confirm existing
beliefs (Hart \& Nisbet, 2012). Unlike coordinated disinformation
campaigns leveraging these vulnerabilities (Garrett, 2017; Nisbet \&
Kamenchuk, 2019), AI hallucinations, though generated without intent,
may also trigger these biases. Currently there is limited understanding
of whether communities resist hallucinated content once it is recognized
-- an example is the resignation of an academic journal editorial board
due to systematic hallucinations from AI editors (Oransky, 2024) -\/- or
whether preexisting biases override potential skepticism.

Besides, intervention at this level also faces a challenge that AI
hallucinations happen without a clear signal of deception originating
from human intent or from traceable agencies. Thus, the meso-level
agenda should focus on developing empirical strategies for tracing
group-level uptake of hallucinations, identifying conditions under which
such content is socially reinforced or rejected, and evaluating whether
existing corrective mechanisms remain effective when dealing with
content lacking clear human authorship or intent. We also have no
evidence on which group is more impacted by which type of hallucination.
For example, citation relevant hallucinations may have a larger impact
on the science community (Alkaissi \& McFarlane, 2023), and a more
narrative and ambivalent hallucinated science story could be more
dangerous for the lay public (Schäfer, 2023). We need an empirically
informed guide on which problems are most acute, for whom, and in what
context, in AI hallucinations for communication.

\textbf{Micro-Level: Cognitive Limitations, and Digital Literacy.} At
the individual level, AI hallucinations may influence users differently
than traditional misinformation because of their stylistic features,
particularly output fluency and apparent confidence. Digital literacy
offers some protection but unevenly distributed among the public. While
users with stronger digital skills are better at identifying
misinformation (Guess et al., 2020; Sirlin et al., 2021). , access alone
does not translate into critical engagement with AI outputs. Younger
users, for example, often misjudge credibility despite high levels of
digital exposure (Menchen-Trevino \& Hargittai, 2011; Wineburg et al.,
2025). Even proficient users fall back on low-effort processing, relying
on peripheral cues like clarity and tone because fluent information
``feels better'' than more complex materials (Petty \& Cacioppo, 1986,
Markowitz, 2024). AI-generated content often matches these preferences
because it appears fluent and authoritative and are readily accessible
(cf. search engine use for ``readily accessible'' information, Hargittai
et al., 2010). AI systems are optimized to perform well on precisely
these cues. Their confident style and rapid delivery encourage shallow
processing, and their design to mirror user preferences can result in
agreeable but inaccurate responses, known as ``sycophancy'' (Sharma et
al., 2023).

Beyond describing this susceptibility to being deceived by AI
hallucinations, understanding the downstream consequences of
encountering AI hallucinations calls for empirical validations. For
example, while current research focused mostly on attitudes towards
(generative) AI the \emph{technology} (e.g., Eom et al., 2024; Greussing
et al., 2025; Yang et al., 2023), further communication research could
explore how exposure to hallucinations influences key outcomes like user
trust in AI-generated information \emph{content} across topics and
subsequent information-seeking behaviors (e.g., verification, continued
AI use, disengagement). Answering these questions is necessary for
developing targeted interventions that move beyond generic digital
literacy and address the specific interpretive challenges posed by
AI-generated falsehoods.

\hypertarget{bridging-levels-and-moving-forward}{%
\subsubsection{Bridging Levels and Moving
Forward}\label{bridging-levels-and-moving-forward}}

Addressing the demand-side challenges of AI hallucinations requires an
integrative approach that connects the macro-, meso- and micro-level
factors as discussed above. Such an approach could draw insights from
diverse fields, including communication, cognitive psychology,
computational methods, and science and technology studies (STS)
(Schäfer, 2023); and adaptive approaches and potentially collaborative
research involving multiple stakeholders to adequately address these
multi-layered challenges within algorithmically shaped information
environments (Krause et al., 2024). From a communication perspective,
studying the demand side of AI hallucinations is essential for
understanding how individuals and groups make sense of information in
environments increasingly populated by AI agents. Research in this area
can clarify the effects of AI-generated content on public understanding,
social interactions, and democratic processes. By identifying
vulnerabilities and effective response strategies, communication
scholarship can contribute directly to fostering more informed and
resilient engagement with AI technologies.

There was such a warning over two decades ago, ``Humans are not
secure\ldots If it {[}a transhuman AI{]} thinks both faster and better
than a human, it can probably take over a human mind through a text-only
terminal.'' (Yudkowsky, 2002) While today's AI systems may not yet reach
transhuman intelligence, the fluency, speed, and persuasive power are
already challenging the stability of human knowledge making processes,
as illustrated by the examples I discussed at the beginning of this
essay. Addressing AI hallucinations involves more than detecting and
correcting falsehoods; it now requires a forward-looking how the
information ecosystem may evolve in response to a new generative
computational agent.

\newpage

\hypertarget{references}{%
\subsection{References}\label{references}}

Akin, H. (2017). \emph{Overview of the Science of Science Communication}
(K. H. Jamieson, D. M. Kahan, \& D. A. Scheufele, Eds.; Vol. 1). Oxford
University Press. https://doi.org/10.1093/oxfordhb/9780190497620.013.3

Alkaissi, H., \& McFarlane, S. I. (2023). Artificial Hallucinations in
ChatGPT: Implications in Scientific Writing. \emph{Cureus},
\emph{15}(2), e35179. https://doi.org/10.7759/cureus.35179

Augenstein, I., Baldwin, T., Cha, M., Chakraborty, T., Ciampaglia, G.
L., Corney, D., DiResta, R., Ferrara, E., Hale, S., Halevy, A., Hovy,
E., Ji, H., Menczer, F., Miguez, R., Nakov, P., Scheufele, D., Sharma,
S., \& Zagni, G. (2024). Factuality challenges in the era of large
language models and opportunities for fact-checking. \emph{Nature
Machine Intelligence}, \emph{6}(8), 852--863.
https://doi.org/10.1038/s42256-024-00881-z

Beets, R. (2024). \emph{A Mixed-Methods Exploration of Publics'
Perceptions of Scientific Uncertainty---ProQuest}.
https://www.proquest.com/docview/3087915935?pq-origsite=gscholar\&fromopenview=true\&sourcetype=Dissertations\%20\&\%20Theses

Bender, E. M., Gebru, T., McMillan-Major, A., \& Shmitchell, S. (2021).
\emph{On the dangers of stochastic parrots: Can language models be too
big?}. 610--623.

Betley, J., Tan, D., Warncke, N., Sztyber-Betley, A., Bao, X., Soto, M.,
Labenz, N., \& Evans, O. (2025). \emph{Emergent Misalignment: Narrow
finetuning can produce broadly misaligned LLMs} (arXiv:2502.17424).
arXiv. https://doi.org/10.48550/arXiv.2502.17424

Bien-Aimé, S., Wu ,Mu, Appelman ,Alyssa, \& and Jia, H. (2025). Who
Wrote It? News Readers' Sensemaking of AI/Human Bylines.
\emph{Communication Reports}, \emph{38}(1), 46--58.
https://doi.org/10.1080/08934215.2024.2424553

Bond, S. (2024, February 8). AI fakes raise election risks as lawmakers
and tech companies scramble to catch up. \emph{NPR}.
https://www.npr.org/2024/02/08/1229641751/ai-deepfakes-election-risks-lawmakers-tech-companies-artificial-intelligence

Brennen, S., Sanderson, Z., \& de la Puerta, C. (2025). \emph{When it
comes to understanding AI's impact on elections, we're still working in
the dark}. Brookings.
https://www.brookings.edu/articles/when-it-comes-to-understanding-ais-impact-on-elections-were-still-working-in-the-dark/

Brumfiel, G. (2021, July 15). The U.S. Surgeon General Is Calling
COVID-19 Misinformation An ``Urgent Threat.'' \emph{NPR}.
https://www.npr.org/sections/health-shots/2021/07/15/1016013826/the-u-s-surgeon-general-is-calling-covid-19-misinformation-an-urgent-threat

Budak, C., Nyhan, B., Rothschild, D. M., Thorson, E., \& Watts, D. J.
(2024). Misunderstanding the harms of online misinformation.
\emph{Nature}, \emph{630}(8015), 45--53.
https://doi.org/10.1038/s41586-024-07417-w

Chen, K., Shao, A., Burapacheep, J., \& Li, Y. (2024). Conversational AI
and equity through assessing GPT-3's communication with diverse social
groups on contentious topics. \emph{Scientific Reports}, \emph{14}(1),
1561. https://doi.org/10.1038/s41598-024-51969-w

Cinelli, M., De Francisci Morales, G., Galeazzi, A., Quattrociocchi, W.,
\& Starnini, M. (2021). The echo chamber effect on social media.
\emph{Proceedings of the National Academy of Sciences}, \emph{118}(9),
e2023301118. https://doi.org/10.1073/pnas.2023301118

Costello, T. H., Pennycook, G., \& Rand, D. G. (2024). Durably reducing
conspiracy beliefs through dialogues with AI. \emph{Science},
\emph{385}(6714), eadq1814. https://doi.org/10.1126/science.adq1814

Crawford, K. (2021). \emph{The atlas of AI: Power, politics, and the
planetary costs of artificial intelligence}.
https://yalebooks.yale.edu/book/9780300264630/atlas-of-ai/

Druckman, J. N., \& Bolsen, T. (2011). Framing, motivated reasoning, and
opinions about emergent technologies. \emph{Journal of Communication},
\emph{61}(4), 659--688. https://doi.org/10.1111/j.1460-2466.2011.01562.x

Dunwoody, S., Hendriks, F., Massarani, L., \& Peters, H. P. (2018).
\emph{How Journalists Deal with Scientific Uncertainty and What That
Means for the Audience}.

Eom, D., Newman, T., Brossard, D., \& Scheufele, D. A. (2024). Societal
guardrails for AI? Perspectives on what we know about public opinion on
artificial intelligence. \emph{Science and Public Policy}, \emph{51}(5),
1004--1013. https://doi.org/10.1093/scipol/scae041

Fan, W., Ding, Y., Ning, L., Wang, S., Li, H., Yin, D., Chua, T.-S., \&
Li, Q. (2024). A Survey on RAG Meeting LLMs: Towards Retrieval-Augmented
Large Language Models. \emph{Proceedings of the 30th ACM SIGKDD
Conference on Knowledge Discovery and Data Mining}, 6491--6501.
https://doi.org/10.1145/3637528.3671470

Garrett, R. K. (2017). The ``echo chamber'' distraction: Disinformation
campaigns are the problem, not audience fragmentation. \emph{Journal of
Applied Research in Memory and Cognition}, \emph{6}(4), 370--376.
https://doi.org/10.1016/j.jarmac.2017.09.011

Golebiewski, M., \& boyd, danah. (2019). \emph{Data Voids: Where Missing
Data Can Easily Be Exploited}.

Greussing, E., Guenther, L., Baram-Tsabari, A., Dabran-Zivan, S., Jonas,
E., Klein-Avraham, I., Taddicken, M., Agergaard, T. E., Beets, B.,
Brossard, D., Chakraborty, A., Fage-Butler, A., Huang, C.-J., Kankaria,
S., Lo, Y.-Y., Nielsen, K. H., Riedlinger, M., \& Song, H. (2025). The
perception and use of generative AI for science-related information
search: Insights from a cross-national study. \emph{Public Understanding
of Science}, 09636625241308493.
https://doi.org/10.1177/09636625241308493

Guess, A. M., Lerner, M., Lyons, B., Montgomery, J. M., Nyhan, B.,
Reifler, J., \& Sircar, N. (2020). A digital media literacy intervention
increases discernment between mainstream and false news in the United
States and India. \emph{Proceedings of the National Academy of
Sciences}, \emph{117}(27), 15536--15545.
https://doi.org/10.1073/pnas.1920498117

Haggart, B. (2023, February 6). \emph{Here's why ChatGPT raises issues
of trust}. World Economic Forum.
https://www.weforum.org/stories/2023/02/why-chatgpt-raises-issues-of-trust-ai-science/

Hargittai, E., Fullerton, L., Menchen-Trevino, E., \& Thomas, K. Y.
(2010). Trust Online: Young Adults' Evaluation of Web Content.
\emph{International Journal of Communication}, \emph{4}(0), Article 0.

Hart, P. S., \& Nisbet, E. C. (2012). Boomerang Effects in Science
Communication: How Motivated Reasoning and Identity Cues Amplify Opinion
Polarization About Climate Mitigation Policies. \emph{Communication
Research}, \emph{39}(6), 701--723.
https://doi.org/10.1177/0093650211416646

Jamieson, K. H., \& Cappella, J. N. (2008). \emph{Echo chamber: Rush
Limbaugh and the conservative media establishment}. Oxford University
Press.

Kaczynski, E. S., Andrew. (2024, January 22). \emph{Fake Joe Biden
robocall urges New Hampshire voters not to vote in Tuesday's Democratic
primary \textbar{} CNN Politics}. CNN.
https://www.cnn.com/2024/01/22/politics/fake-joe-biden-robocall/index.html

Kidman, A. (2025, February 23). \emph{No, your computer isn't powered by
tiny microscopic bees}. Alex Reviews Tech.
https://alexreviewstech.com/no-your-computer-isnt-powered-by-tiny-microscopic-bees/

Kington, R. S., Arnesen, S., Chou, W.-Y. S., Curry, S. J., Lazer, D., \&
Villarruel, A. M. (2021). Identifying Credible Sources of Health
Information in Social Media: Principles and Attributes. \emph{NAM
Perspectives}, 10.31478/202107a. https://doi.org/10.31478/202107a

Kirk, H. R., Vidgen, B., Röttger, P., \& Hale, S. A. (2024). The
benefits, risks and bounds of personalizing the alignment of large
language models to individuals. \emph{Nature Machine Intelligence},
\emph{6}(4), 383--392. https://doi.org/10.1038/s42256-024-00820-y

Krause, N. M., Freiling, I., \& Scheufele, D. A. (2022). The
``Infodemic'' Infodemic: Toward a More Nuanced Understanding of
Truth-Claims and the Need for (Not) Combatting Misinformation. \emph{The
ANNALS of the American Academy of Political and Social Science},
\emph{700}(1), 112--123. https://doi.org/10.1177/00027162221086263

Krause, N. M., Freiling, I., \& Scheufele, D. A. (2024). \emph{Our
changing information ecosystem for science and why it matters for
effective science communication}. OSF.
https://doi.org/10.31235/osf.io/j9yh7

Latour, B. (2005). \emph{Reassembling the Social: An Introduction to
Actor-Network-Theory}. OUP Oxford.

Lewandowsky, S., Ecker, U. K. H., Seifert, C. M., Schwarz, N., \& Cook,
J. (2012). Misinformation and Its Correction: Continued Influence and
Successful Debiasing. \emph{Psychological Science in the Public
Interest, Supplement}, \emph{13}(3), 106--131.
https://doi.org/10.1177/1529100612451018

Li, F., Yang, Y., \& Yu, G. (2025). Nudging Perceived Credibility: The
Impact of AIGC Labeling on User Distinction of AI-Generated Content.
\emph{Emerging Media}, 27523543251317572.
https://doi.org/10.1177/27523543251317572

Liao, Q. V., \& Vaughan, J. W. (2024). AI Transparency in the Age of
LLMs: A Human-Centered Research Roadmap. \emph{Harvard Data Science
Review}, \emph{Special Issue 5}.
https://doi.org/10.1162/99608f92.8036d03b

Loukissas, Y. A. (2019). \emph{All Data Are Local: Thinking Critically
in a Data-Driven Society}. MIT Press.

Lu, T. (2025). \emph{Maximum Hallucination Standards for Domain-Specific
Large Language Models} (arXiv:2503.05481). arXiv.
https://doi.org/10.48550/arXiv.2503.05481

Markowitz, D. M. (2024). From complexity to clarity: How AI enhances
perceptions of scientists and the public's understanding of science.
\emph{PNAS Nexus}, \emph{3}(9), pgae387.
https://doi.org/10.1093/pnasnexus/pgae387

Menchen-Trevino, E., \& Hargittai, E. (2011). Young adults' credibility
assessment of Wikipedia. \emph{Information Communication and Society},
\emph{14}(1), 24--51. https://doi.org/10.1080/13691181003695173

Moffatt v. Air Canada, SC-2023-005609 (Civil Resolution Tribunal of
British Columbia February 14, 2024). https://canlii.ca/t/k2spq

Montasari, R. (2024). The Dual Role of Artificial Intelligence in Online
Disinformation: A Critical Analysis. In R. Montasari (Ed.),
\emph{Cyberspace, Cyberterrorism and the International Security in the
Fourth Industrial Revolution: Threats, Assessment and Responses} (pp.
229--240). Springer International Publishing.
https://doi.org/10.1007/978-3-031-50454-9\_11

Narayanan Venkit, P., Chakravorti, T., Gupta, V., Biggs, H., Srinath,
M., Goswami, K., Rajtmajer, S., \& Wilson, S. (2024). An Audit on the
Perspectives and Challenges of Hallucinations in NLP. In Y. Al-Onaizan,
M. Bansal, \& Y.-N. Chen (Eds.), \emph{Proceedings of the 2024
Conference on Empirical Methods in Natural Language Processing} (pp.
6528--6548). Association for Computational Linguistics.
https://doi.org/10.18653/v1/2024.emnlp-main.375

National Academies of Sciences, Engineering, and Medicine. (2024).
\emph{Understanding and Addressing Misinformation about Science
\textbar{} National Academies}.
https://www.nationalacademies.org/our-work/understanding-and-addressing-misinformation-about-science

Nisbet, E. C., \& Kamenchuk, O. (2019). The Psychology of
State-Sponsored Disinformation Campaigns and Implications for Public
Diplomacy. \emph{Hague Journal of Diplomacy}, \emph{14}(1--2), 65--82.
https://doi.org/10.1163/1871191x-11411019

Omiye, J. A., Lester, J. C., Spichak, S., Rotemberg, V., \& Daneshjou,
R. (2023). Large language models propagate race-based medicine.
\emph{Npj Digital Medicine}, \emph{6}(1), 1--4.
https://doi.org/10.1038/s41746-023-00939-z

Oransky, I. (2024, December 27). Evolution journal editors resign en
masse to protest Elsevier changes. \emph{Retraction Watch}.
https://retractionwatch.com/2024/12/27/evolution-journal-editors-resign-en-masse-to-protest-elsevier-changes/

Owen, L. (2025, March 4). The L.A. Times adds AI-generated counterpoints
to its opinion pieces and guess what, there are problems. \emph{Nieman
Lab}.
https://www.niemanlab.org/2025/03/the-l-a-times-adds-ai-generated-counterpoints-to-its-opinion-pieces-and-guess-what-there-are-problems/

Ozmen Garibay, O., Winslow, B., Andolina, S., Antona, M., Bodenschatz,
A., Coursaris, C., Falco, G., Fiore, S. M., Garibay, I., Grieman, K.,
Havens, J. C., Jirotka, M., Kacorri, H., Karwowski, W., Kider, J.,
Konstan, J., Koon, S., Lopez-Gonzalez, M., Maifeld-Carucci, I., \&
McGregor, S. (2023). Six Human-Centered Artificial Intelligence Grand
Challenges. \emph{International Journal of Human-Computer Interaction},
\emph{39}(3), 391--437. https://doi.org/10.1080/10447318.2022.2153320

Palta, R., Angwin, J., \& Nelson, A. (2024, February 27). \emph{How We
Tested Leading AI Models Performance on Election Queries}. Proof.
https://www.proofnews.org/how-we-tested-leading-ai-models-performance-on-election-queries/

Pawitan, Y., \& Holmes, C. (2025). Confidence in the Reasoning of Large
Language Models. \emph{Harvard Data Science Review}, \emph{7}(1).
https://doi.org/10.1162/99608f92.b033a087

Pennycook, G., Epstein, Z., Mosleh, M., Arechar, A. A., Eckles, D., \&
Rand, D. G. (2021). Shifting attention to accuracy can reduce
misinformation online. \emph{Nature}, \emph{592}(7855), Article 7855.
https://doi.org/10.1038/s41586-021-03344-2

Peters, H. P., \& Dunwoody, S. (2016). Scientific uncertainty in media
content: Introduction to this special issue. \emph{Public Understanding
of Science}, \emph{25}(8), 893--908.
https://doi.org/10.1177/0963662516670765

Rammert, W. (2008). \emph{Where the action is: Distributed agency
between humans, machines, and programs}.

Reid, L. (2024, May 14). \emph{Generative AI in Search: Let Google do
the searching for you}. Google.
https://blog.google/products/search/generative-ai-google-search-may-2024/

Reilly L. (2025, March 5). \emph{The LA Times' new AI tool sympathized
with the KKK. Its owner wasn't aware until hours later \textbar{} CNN
Business}. CNN.
https://www.cnn.com/2025/03/05/media/la-times-ai-kkk-comments/index.html

Schäfer, M. S. (2016). Mediated trust in science: Concept, measurement
and perspectives for the `science of science communication'.
\emph{Journal of Science Communication}, \emph{15}(5), C02.
https://doi.org/10.22323/2.15050302

Schäfer, M. S. (2023). The Notorious GPT: Science communication in the
age of artificial intelligence. \emph{Journal of Science Communication},
\emph{22}(2), Y02. https://doi.org/10.22323/2.22020402

Scheufele, D. A. (2014). Science communication as political
communication. \emph{Proceedings of the National Academy of Sciences},
\emph{111}(Supplement 4), 13585--13592.
https://doi.org/10.1073/pnas.1317516111

Scheufele, D. A., Hoffman, A. J., Neeley, L., \& Reid, C. M. (2021).
Misinformation about science in the public sphere. \emph{Proceedings of
the National Academy of Sciences}, \emph{118}(15).

Scheufele, D. A., \& Krause, N. M. (2019). Science audiences,
misinformation, and fake news. \emph{Proceedings of the National Academy
of Sciences of the United States of America}, \emph{116}(16),
7662--7669. https://doi.org/10.1073/pnas.1805871115

Scheufele, D. A., Krause, N. M., \& Freiling, I. (2021). Misinformed
about the ``infodemic?'' Science's ongoing struggle with misinformation.
\emph{Journal of Applied Research in Memory and Cognition},
\emph{10}(4), 522--526. https://doi.org/10.1016/j.jarmac.2021.10.009

Sharma, M., Tong, M., Korbak, T., Duvenaud, D., Askell, A., Bowman, S.
R., Durmus, E., Hatfield-Dodds, Z., Johnston, S. R., Kravec, S. M.,
Maxwell, T., McCandlish, S., Ndousse, K., Rausch, O., Schiefer, N., Yan,
D., Zhang, M., \& Perez, E. (2023, October 13). \emph{Towards
Understanding Sycophancy in Language Models}. The Twelfth International
Conference on Learning Representations.
https://openreview.net/forum?id=tvhaxkMKAn

Shepardson, D. (2024, September 26). Consultant fined \$6 million for
using AI to fake Biden's voice in robocalls. \emph{Reuters}.
https://www.reuters.com/world/us/fcc-finalizes-6-million-fine-over-ai-generated-biden-robocalls-2024-09-26/

Shumailov, I., Shumaylov, Z., Zhao, Y., Papernot, N., Anderson, R., \&
Gal, Y. (2024). AI models collapse when trained on recursively generated
data. \emph{Nature}, \emph{631}, 755--759.
https://doi.org/10.1038/s41586-024-07566-y

Sirlin, N., Epstein, Z., Arechar, A. A., \& Rand, D. G. (2021). Digital
literacy is associated with more discerning accuracy judgments but not
sharing intentions. \emph{Harvard Kennedy School Misinformation Review}.
https://doi.org/10.37016/mr-2020-83

Tewari, S., Zabounidis, R., Kothari, A., Bailey, R., \& Alm, C. O.
(2021). Perceptions of Human and Machine-Generated Articles.
\emph{Digital Threats: Research and Practice}, \emph{2}(2), 1--16.
https://doi.org/10.1145/3428158

Thirunavukarasu, A. J., Ting, D. S. J., Elangovan, K., Gutierrez, L.,
Tan, T. F., \& Ting, D. S. W. (2023). Large language models in medicine.
\emph{Nature Medicine}, 1--11.

Thorbecke, C., \& Duffy, C. (2024, February 22). \emph{Google halts AI
tool's ability to produce images of people after backlash \textbar{} CNN
Business}. CNN.
https://www.cnn.com/2024/02/22/tech/google-gemini-ai-image-generator/index.html

Tian, S., Jin, Q., Yeganova, L., Lai, P.-T., Zhu, Q., Chen, X., Yang,
Y., Chen, Q., Kim, W., Comeau, D. C., Islamaj, R., Kapoor, A., Gao, X.,
\& Lu, Z. (2023). Opportunities and challenges for ChatGPT and large
language models in biomedicine and health. \emph{Briefings in
Bioinformatics}, \emph{25}(1), bbad493.
https://doi.org/10.1093/bib/bbad493

van der Linden, S. (2023). \emph{Foolproof: Why Misinformation Infects
Our Minds and How to Build Immunity}. W. W. Norton \& Company.

Vectara. (2024). \emph{Hallucination Leaderboard}.
https://github.com/vectara/hallucination-leaderboard

Voigt, P., \& von dem Bussche, A. (2017). The EU General Data Protection
Regulation (GDPR): A Practical Guide. \emph{Springer Nature eBook}.
https://doi.org/10.1007/978-3-319-57959-7

Weidinger, L., Mellor, J., Rauh, M., Griffin, C., Uesato, J., Huang,
P.-S., Cheng, M., Glaese, M., Balle, B., Kasirzadeh, A., Kenton, Z.,
Brown, S., Hawkins, W., Stepleton, T., Biles, C., Birhane, A., Haas, J.,
Rimell, L., Hendricks, L. A., \ldots{} Gabriel, I. (2021). \emph{Ethical
and social risks of harm from Language Models} (arXiv:2112.04359).
arXiv. https://doi.org/10.48550/arXiv.2112.04359

Westerlund, M. (2019). The emergence of deepfake technology: A review.
\emph{Technology Innovation Management Review}, \emph{9}(11).

Wineburg, S., McGrew, S., Breakstone, J., \& Ortega, T. (2025).
\emph{Evaluating Information: The Cornerstone of Civic Online
Reasoning}. https://purl.stanford.edu/fv751yt5934

Wood, M. C., \& Forbes, A. A. (2024). \emph{100\% Hallucination
Elimination Using Acurai} (arXiv:2412.05223). arXiv.
https://doi.org/10.48550/arXiv.2412.05223

World Health Organization. (2020). \emph{Managing the COVID-19
infodemic: Promoting healthy behaviours and mitigating the harm from
misinformation and disinformation}.
https://www.who.int/news/item/23-09-2020-managing-the-covid-19-infodemic-promoting-healthy-behaviours-and-mitigating-the-harm-from-misinformation-and-disinformation

Xiong, M., Hu, Z., Lu, X., Li, Y., Fu, J., He, J., \& Hooi, B. (2023).
Can LLMs Express Their Uncertainty? An Empirical Evaluation of
Confidence Elicitation in LLMs. \emph{arXiv Preprint arXiv:2306.13063}.

Xu, Z., Jain, S., \& Kankanhalli, M. (2025). \emph{Hallucination is
Inevitable: An Innate Limitation of Large Language Models}
(arXiv:2401.11817). arXiv. https://doi.org/10.48550/arXiv.2401.11817

Yang, S., Krause, N. M., Bao, L., Calice, M. N., Newman, T. P.,
Scheufele, D. A., Xenos, M. A., \& Brossard, D. (2023). In AI We Trust:
The Interplay of Media Use, Political Ideology, and Trust in Shaping
Emerging AI Attitudes. \emph{Journalism \& Mass Communication
Quarterly}, 10776990231190868. https://doi.org/10.1177/10776990231190868

Yudkowsky, E. (2002). \emph{The AI-Box Experiment}.
https://www.yudkowsky.net/singularity/aibox

Zamfirescu-Pereira, J. D., Wong, R. Y., Hartmann, B., \& Yang, Q.
(2023). Why Johnny Can't Prompt: How Non-AI Experts Try (and Fail) to
Design LLM Prompts. \emph{Proceedings of the 2023 CHI Conference on
Human Factors in Computing Systems}, 1--21.
https://doi.org/10.1145/3544548.3581388

Zhang, Y., Li, Y., Cui, L., Cai, D., Liu, L., Fu, T., Huang, X., Zhao,
E., Zhang, Y., Chen, Y., Wang, L., Luu, A. T., Bi, W., Shi, F., \& Shi,
S. (2023). \emph{Siren's Song in the AI Ocean: A Survey on Hallucination
in Large Language Models} (arXiv:2309.01219). arXiv.
https://doi.org/10.48550/arXiv.2309.01219

Zhao, G. (2024, December 4). Stanford misinformation expert admits to
ChatGPT ``hallucinations'' in court statement. \emph{The Stanford
Daily}. https://stanforddaily.com/2024/12/04/hancock-admitted-to-ai-use/

Zou, W., Geng, R., Wang, B., \& Jia, J. (2024). \emph{PoisonedRAG:
Knowledge Corruption Attacks to Retrieval-Augmented Generation of Large
Language Models} (arXiv:2402.07867). arXiv.
https://doi.org/10.48550/arXiv.2402.07867

\end{document}